\documentclass[aip,apl,preprint]{revtex4-2}
\usepackage{amsmath}
\usepackage{graphicx}
\usepackage{booktabs}
\usepackage{hyperref}
\hypersetup{hidelinks}
\graphicspath{{./figures/}}

\begin{document}

\title{Above-bulk DC Kerr electro-optics at the water/ITO interface, resolved with the Pockels effect}

\author{Soichiro Ashikaga}
\affiliation{Department of Physics, Tokyo University of Science, 1-3 Kagurazaka, Shinjuku-ku, Tokyo 162-8601, Japan}

\author{Kazuaki Nakata}
\affiliation{Department of Physics, Tokyo University of Science, 1-3 Kagurazaka, Shinjuku-ku, Tokyo 162-8601, Japan}

\author{Akihiro Okada}
\affiliation{Department of Physics, Tokyo University of Science, 1-3 Kagurazaka, Shinjuku-ku, Tokyo 162-8601, Japan}

\author{Takumi Takayanagi}
\affiliation{Department of Physics, Tokyo University of Science, 1-3 Kagurazaka, Shinjuku-ku, Tokyo 162-8601, Japan}

\author{Kyohei Yamashita}
\affiliation{Department of Physics, Tokyo University of Science, 1-3 Kagurazaka, Shinjuku-ku, Tokyo 162-8601, Japan}

\author{Takayoshi Kobayashi}
\affiliation{National Yang Ming Chiao Tung University, No. 1001, Daxue Road, East District, Hsinchu City 300093, Taiwan}

\author{Eiji Tokunaga}
\email{eiji@rs.tus.ac.jp}
\affiliation{Department of Physics, Tokyo University of Science, 1-3 Kagurazaka, Shinjuku-ku, Tokyo 162-8601, Japan}

\begin{abstract}
On a charged interface, broken inversion symmetry permits a large field-linear Pockels response through $\chi^{(2)}(\omega;\omega,0)$; at the water/ITO interface $|r_{13}|$ has been reported to reach the $10^{2}\,\mathrm{pm/V}$ order.
The coexisting third-order DC Kerr term $\chi^{(3)}(\omega;\omega,0,0)$---small in bulk water ($|\chi^{(3)}_{\mathrm{bulk}}|\sim5.5\times10^{-21}\,\mathrm{m^2/V^2}$)---had not been jointly parameterized with the Pockels term along the fundamental-frequency ($\omega$) electro-optic path.
Superimposing an AC modulation and a DC bias in 0.1\,M NaCl mixes the $\chi^{(3)}$ contribution with the 1f response through the cross-term $2 s_{1133}\,E_{\mathrm{DC}}$, so that the water refractive-index modulation $\Delta n_{\mathrm{water}}$ varies linearly with $V_{\mathrm{WE}}$; a model-assisted linear fit then determines both terms from a single AC$+$DC sweep.
At $V_{\mathrm{WE}}=0\,\mathrm{V}$ vs Ag/AgCl, $|r_{13}|=(1.18 \pm 0.06_{\mathrm{PZC}})\times10^{2}\,\mathrm{pm/V}$ and, under $\eta_{\mathrm{DC}}=1$, the thickness-normalized DC Kerr coefficient $|s_{1133}/d_{\mathrm{EDL}}|=33.0 \pm 5.6\,\mathrm{pm/V^{2}}$.
Across physically reasonable $d_{\mathrm{EDL}}$ ($0.6$--$1.6\,\mathrm{nm}$), the interfacial DC Kerr susceptibility reaches $|\chi^{(3),\mathrm{int}}_{1133}| \approx (2\text{--}5.5)\times10^{-20}\,\mathrm{m^2/V^2}$, several-fold above the visible-range bulk-water value.
This response is a property of the specific interface, tunable through the choice of electrode, electrolyte, and solvent rather than intrinsic to bulk water.
Amid renewed interest in the Kerr response of water (including recent THz-band optical Kerr studies), the method directly probes this DC Kerr term along the $\omega$ path and complements SHG/SFG (the $2\omega$ path).
\end{abstract}
\maketitle
In bulk water, a centrosymmetric isotropic liquid, inversion symmetry forbids
the second-order susceptibility $\chi^{(2)}$ while leaving the third-order
susceptibility $\chi^{(3)}$ allowed. The DC Kerr response
$\chi^{(3)}(\omega;\omega,0,0)$ remains of order
$|\chi^{(3)}_{\mathrm{bulk}}|\sim 5.5\times10^{-21}\,\mathrm{m^2/V^2}$ in the
visible range~\cite{zahn1985electro,aroney1976kerr}. The THz-band optical
Kerr effect has recently drawn renewed attention to the Kerr response of
water~\cite{tcypkin2019high,zalden2018molecular}.

At a charged interface, broken inversion symmetry permits a field-linear EO response through the Pockels effect mediated by $\chi^{(2)}(\omega;\omega,0)$.
At the water/ITO interface, $|r_{13}|$ has been repeatedly reported to reach the $10^{2}\,\mathrm{pm/V}$ order~\cite{tokunaga2007pockels,okada2022interfacial}.
A comparably large Pockels response ($|r_{13}|\sim10^{2}\,\mathrm{pm/V}$) has also been reported for water in contact with a TiO$_2$-covered Ti electrode~\cite{tanimoto2023pockels}, indicating that this regime is not specific to the water/ITO interface.
A proposed mechanism attributes this response to a dynamic expansion of the electric double layer (EDL) under AC field application that breaks inversion symmetry within bulk water~\cite{yukita2018mechanisms}.
In contrast, the coexisting $\chi^{(3)}(\omega;\omega,0,0)$-mediated field-quadratic EO response (the interfacial DC Kerr term) gives a much smaller signal at the conventional 2f Kerr frequency~\cite{tokunaga2007pockels}, and no model-assisted framework has jointly parameterized it with the Pockels term along the fundamental-frequency $\omega$ EO path.
A related separation is well established in $2\omega$ SHG/SFG, where phase-sensitive measurements resolve $\chi^{(2)}_{\mathrm{eff}} = \chi^{(2)}_{\mathrm{s}} + \chi^{(3)}_{\mathrm{b}}\Phi_0$ into the interface-specific $\chi^{(2)}_{\mathrm{s}}$ response and the $\chi^{(3)}_{\mathrm{b}}\Phi_0$ response induced in the EDL and bulk water by the interfacial electrostatic potential $\Phi_0$~\cite{ohno2016phase,dalstein2019direct}.
This separation, however, operates along the $2\omega$ path and is not equivalent to the fundamental-frequency EO problem in either the frequency argument ($2\omega$ vs $\omega$) or the DC-field input ($\Phi_0$ vs $E_{\mathrm{DC}}$).

Taking the SHG/SFG separation as conceptual motivation, here we exploit the fact that a DC bias on the working-electrode potential $V_{\mathrm{WE}}$ (referenced to Ag/AgCl) mixes the $\chi^{(3)}$ contribution into the fundamental-frequency ($\omega$) 1f response through the cross-term $2 s_{1133}\,E_{\mathrm{DC}}$, so that the refractive-index change of water at 1f varies linearly with $V_{\mathrm{WE}}$ and the Pockels and DC Kerr terms can be jointly parameterized from a single AC$+$DC sweep.
Combining the systematic $V_{\mathrm{WE}}$ sweep with $\Delta L_{\mathrm{opt}}/R_p$ normalization, a single weighted least-squares (WLS) fit maps $r_{13}$ and $s_{1133}/d_{\mathrm{EDL}}$ onto separate coefficients, from which we extract the effective interfacial DC Kerr electro-optic coefficient $|s_{1133}/d_{\mathrm{EDL}}|$.
This extends prior Pockels-only extraction routes~\cite{tokunaga2007pockels,okada2022interfacial} to an applied-EO path that acquires both the Pockels and DC Kerr components at a liquid-electrolyte interface in a single measurement.
We find the interfacial DC Kerr susceptibility to lie several-fold above its bulk-water value, marking it as a tunable property of the charged interface rather than an invariant of bulk water.

The water/ITO interface possesses rotational symmetry about the surface normal ($C_{\infty v}$)~\cite{eisenthal1996liquid},
and the probe light is unpolarized and normally incident, so the optical field lies entirely in the interface ($x$–$y$) plane.
The measurement then probes the change in the in-plane refractive index produced by a DC field $E_{3}$ along the surface
normal ($z$ axis, component 3), so the extracted quantities are the effective electro-optic coefficients $r_{13}$ and
$s_{1133}$ that couple this normal field to the in-plane probe polarization.
From Boyd's electro-optic indicatrix expansion~\cite{boyd2020} (a power series in $E$,
$\eta_{ij} = \eta^{(0)}_{ij} + r_{ijk} E_k + s_{ijkl} E_k E_l + \cdots$), via the variation
$\Delta n_{\mathrm{water}} = -(n^3/2)\,\Delta\eta$ of $\eta = 1/n^2$, the refractive-index change is
\begin{equation}
  \Delta n_{\mathrm{water}} = -\frac{n^3}{2}\bigl(r_{13}\,E + s_{1133}\,E^2\bigr).
  \label{eq:dn_rs}
\end{equation}
Here $r_{13}$ is the electro-optic coefficient associated with $\chi^{(2)}(\omega;\omega,0)$
and $s_{1133}$ with $\chi^{(3)}(\omega;\omega,0,0)$.
When a DC component $E_{\mathrm{DC}}$ and an AC component $E_{\mathrm{AC}}\cos(\omega t)$ are superimposed,
the peak amplitude of the 1f component is
$\Delta n_{\mathrm{water},1f} = -(n^3/2)\,E_{\mathrm{AC}}\,(r_{13} + 2 s_{1133}\,E_{\mathrm{DC}})$, where the second term
$2 s_{1133}\,E_{\mathrm{DC}}$ acts as the cross-term that mixes the $\chi^{(3)}$ contribution into the
fundamental-frequency response.
A purely $\chi^{(3)}$ induced response also appears along the 2f path at order
$\propto s_{1133}\,E_{\mathrm{AC}}^2\cos(2\omega t)$; however, at this 2f path, where the Kerr effect would
conventionally be measured, the water/ITO interface has been reported to give a much smaller signal~\cite{tokunaga2007pockels}.
The 1f bias-mixing path used here therefore provides an alternative route that constrains both components
simultaneously through the more sensitive fundamental-frequency response.

The interfacial optical path-length change within the EDL that the multilayer transfer-matrix fit directly
constrains is $\Delta L_{\mathrm{opt}} = d_{\mathrm{EDL}}\,\Delta n_{\mathrm{water}}$.
Normalizing this by the AC peak voltage $R_p$ measured at the reference electrode gives
$\Delta L_{\mathrm{opt}}/R_p = -(\eta_{\mathrm{AC}}\,n^3/2)(r_{13} + 2 s_{1133}\,E_{\mathrm{DC}})$,
where $\eta_{\mathrm{AC}}$ is the AC voltage amplitude partitioning ratio, the fraction of the measured AC peak voltage $R_p$ that drops across the EDL.
Substituting the DC field inside the EDL, $E_{\mathrm{DC}} = \eta_{\mathrm{DC}}\,V_{\mathrm{WE}}/d_{\mathrm{EDL}}$,
where $\eta_{\mathrm{DC}}$ is the DC voltage amplitude partitioning ratio,
the observable becomes a linear function of $V_{\mathrm{WE}}$,
\begin{equation}
  \frac{\Delta L_{\mathrm{opt}}}{R_p}(V_{\mathrm{WE}}) = L_1 + L_2\,V_{\mathrm{WE}},
  \label{eq:fit_equation}
\end{equation}
with coefficients
$L_1 = -(\eta_{\mathrm{AC}}\,n^3/2)\,r_{13}$ and
$L_2 = -\eta_{\mathrm{AC}}\,\eta_{\mathrm{DC}}\,n^3\,(s_{1133}/d_{\mathrm{EDL}})$
corresponding to $r_{13}$ and $s_{1133}/d_{\mathrm{EDL}}$, respectively.
The mapping from $L_2$ to $s_{1133}/d_{\mathrm{EDL}}$ needs no value of $d_{\mathrm{EDL}}$, so this
thickness-normalized quantity is reported directly; recovering the absolute $|\chi^{(3)}_{1133}|$ does
require a separate $d_{\mathrm{EDL}}$ assumption.
The stability of $\eta_{\mathrm{AC}}$ against $V_{\mathrm{WE}}$ is experimentally bounded by
DC-bias-dependent electrochemical impedance spectroscopy (EIS) ($\pm 1\%$; see the supplementary material), whereas the bias dependence of the optical
transfer response, together with the operational value of $\eta_{\mathrm{DC}}$, remain model assumptions.
The EDL field profiles $E_{\mathrm{AC}}(z)$ and $E_{\mathrm{DC}}(z)$ are treated here as a uniform-field
layer; the general depth-weighted form of the slope and its reduction to this expression are given in the supplementary material.

\begin{figure}[tbp]
  \centering
  \includegraphics[width=\linewidth]{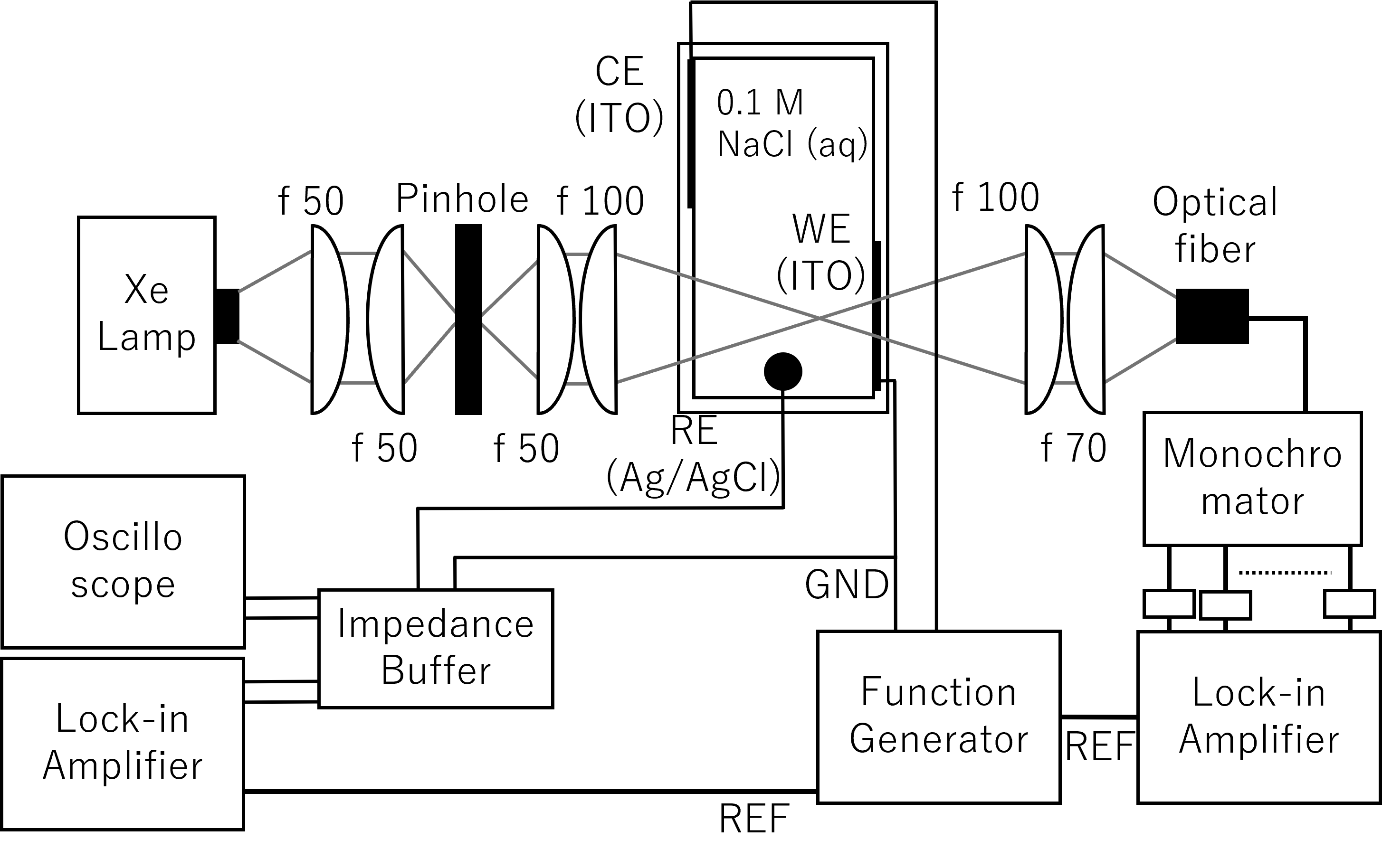}
  \caption{
    Schematic of the electrochemical cell and optical system.
    The cell consists of an ITO transparent electrode (300\,nm thick, on a quartz substrate), 0.1\,M NaCl aqueous solution, an Ag/AgCl reference electrode, and a unity-gain impedance buffer.
    A light source (Xe lamp) is normally incident along the $z$ axis (the interface normal), and the light transmitted through the electric double layer (EDL) at the ITO/water interface is detected with a 128-channel lock-in amplifier (referenced to 21\,Hz).
    An AC modulation (21\,Hz, four $V_{\mathrm{pp}}$ sets) is superimposed on the DC bias and applied across the cell, with the WE held at ground; the AC 1f component (peak voltage $R_p$) and the DC component $V_{\mathrm{WE}}$ are measured simultaneously through the impedance buffer.
  }
  \label{fig:setup}
\end{figure}

A Xe lamp was used as the light source; after collimation, the beam was focused near the water/ITO interface, with its axis along the surface normal.
The transmitted light was detected with a 128-channel lock-in amplifier (21\,Hz reference) and recorded in parallel as the electromodulation spectrum $\Delta T/T$~\cite{tokunaga2007pockels,okada2022interfacial} (Fig.~\ref{fig:setup}).
A multilayer transfer-matrix model comprising the ITO, interfacial EDL, and aqueous-solution layers was fitted to the measured $\Delta T/T$ spectra, and the optical path-length change $\Delta L_{\mathrm{opt}}$ within the EDL layer was extracted.
The electrochemical cell was a three-electrode configuration in which an ITO transparent electrode (300\,nm thick, on a quartz substrate) served as the working electrode (WE) in 0.1\,M aqueous NaCl, with a counter electrode (CE) of the same ITO type placed 17\,mm away in a parallel-plate arrangement (see the supplementary material).
The output of an Ag/AgCl reference electrode inserted into the bulk solution was read out through a unity-gain impedance buffer.
The probe light passed once through the WE side only (transmission order: water $\to$ ITO $\to$ quartz substrate); the CE-side ITO was not in the optical path.
Measurements were grouped by AC amplitude into four sets, $V_{\mathrm{pp}} =$ 0.25, 0.50, 0.75, and 1.00\,V.
For each set, the DC offset $V_{\mathrm{DC}}$ of the function-generator output was stepped under the peak-voltage constraint $|V_{\mathrm{DC}} + V_{\mathrm{pp}}/2| \le 1$\,V, imposed to avoid driving electrochemical reactions at the electrodes; this limit sets the accessible DC range at each amplitude and gives the amplitude-dependent, asymmetric sweep (33 points across the four sets).
With the WE held at ground, the electrode potential $V_{\mathrm{WE}}$---read against the Ag/AgCl reference through the unity-gain buffer---follows the open-circuit potential plus the WE-side share of the applied voltage (roughly half, the rest dropping at the CE); the constraint therefore applies to the generator output, not directly to $V_{\mathrm{WE}}$. $V_{\mathrm{WE}}$ is the independent variable of the fit, not a feedback control, and its recorded range includes the vicinity of $V_{\mathrm{WE}} = 0$.
Details of the optical system, polarization configuration, electrode dimensions, and drive conditions are given in the supplementary material.

Following the approach of Okada \textit{et al.}~\cite{okada2022interfacial}, we determine the AC voltage amplitude partitioning for our own cell by fitting a three-block series equivalent circuit (described in the supplementary material) to broadband EIS, giving $\eta_{\mathrm{AC}} = 0.99$ at 21\,Hz.
The DC partitioning $\eta_{\mathrm{DC}}$ is not measured directly, but down to the sub-hertz lower limit of the EIS the AC voltage already drops almost entirely across the EDL, so we take $\eta_{\mathrm{DC}} = 1$ as the corresponding DC limit.
Since $\eta_{\mathrm{DC}} \le 1$, a smaller value would scale $|s_{1133}/d_{\mathrm{EDL}}|$ upward by $1/\eta_{\mathrm{DC}}$, and the value reported here is therefore a conditional calibration under $\eta_{\mathrm{DC}} = 1$.

\begin{figure}[tbp]
  \centering
  \includegraphics[width=0.85\linewidth]{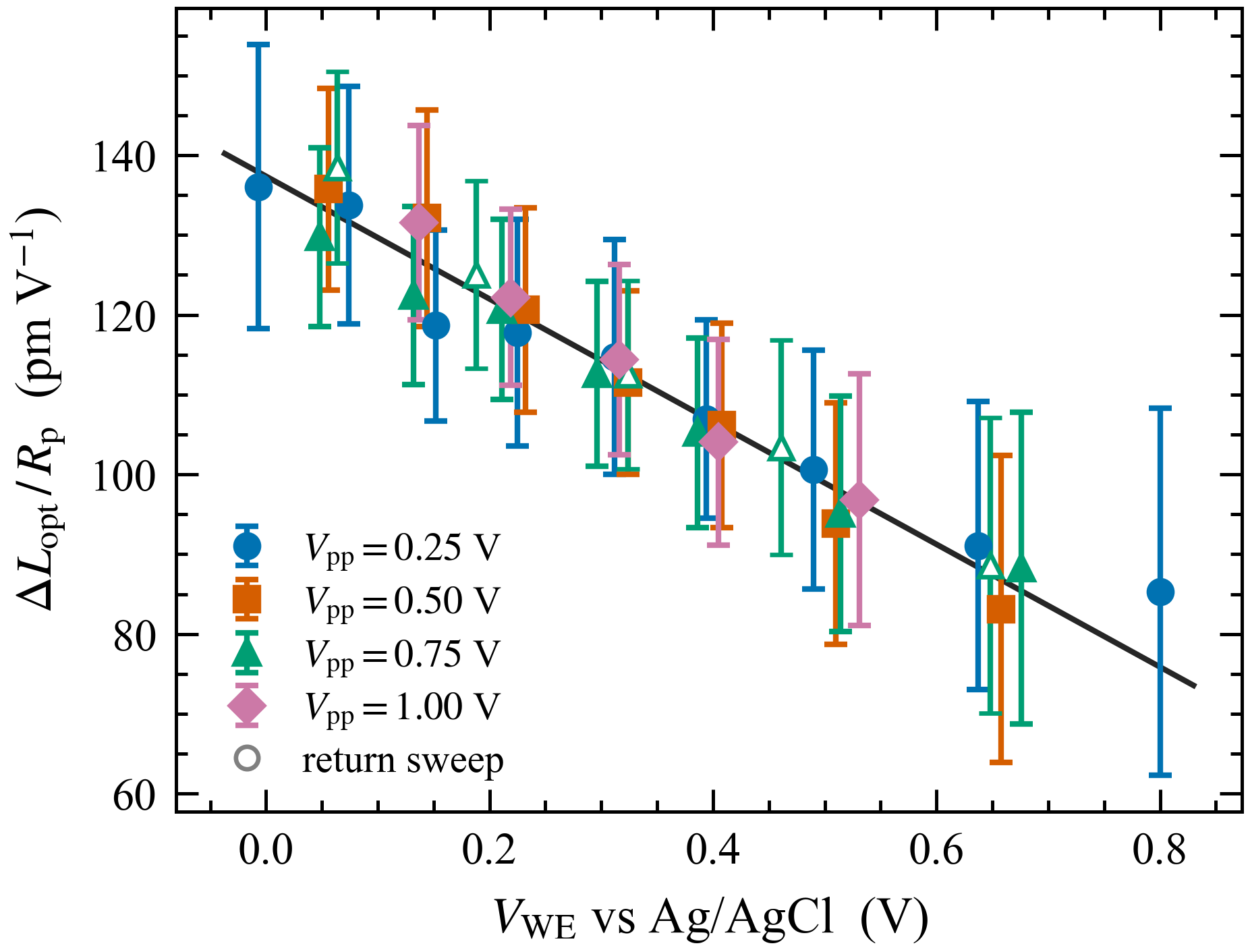}
  \caption{
    DC-bias ($V_{\mathrm{WE}}$) dependence of the primary observable $\Delta L_{\mathrm{opt}}/R_p$ (PZC-uncorrected).
    The four $V_{\mathrm{pp}}$ groups (0.25, 0.50, 0.75, and 1.00\,V) are distinguished by different symbols, and each point carries a statistical error bar from the multilayer transfer-matrix fit.
    The weighted least-squares fit line (solid) corresponds to Eq.~\eqref{eq:fit_equation} and gives the intercept $L_1$ and slope $L_2$.
    The five reverse-sweep points included at $V_{\mathrm{pp}} = 0.75$\,V agree with the forward-sweep points within error.
  }
  \label{fig:hero}
\end{figure}

Figure~\ref{fig:hero} shows $\Delta L_{\mathrm{opt}}$, obtained from the multilayer transfer-matrix fit and normalized by the measured $R_p$, as a function of $V_{\mathrm{WE}}$.
The four $V_{\mathrm{pp}}$ groups (0.25, 0.50, 0.75, and 1.00\,V; 33 points in total) collapse onto a single line within error, confirming that the $1/R_p$ normalization eliminates the AC-amplitude dependence.
A weighted least-squares fit to Eq.~\eqref{eq:fit_equation} gives
\begin{align}
  L_1 &= 137.4 \pm 4.5\;\text{pm/V}, \label{eq:L1_value}\\
  L_2 &= -76.9 \pm 13.1\;\text{pm/V}^2. \label{eq:L2_value}
\end{align}
The uncertainty in $L_1$ includes, in addition to the statistical component of the weighted least-squares fit, a systematic component related to the point of zero charge (PZC), propagated from the session-to-session scatter of the $V_{\mathrm{WE}}$ reference point across days, $\sigma_s \approx 0.09$\,V (supplementary material).
The latter arises because the PZC is not measured, and propagates directly into $|r_{13}|$ derived from $L_1$; reported values for the same water/ITO system in 0.1\,M NaCl nonetheless place the PZC close to our $V_{\mathrm{WE}} = 0$ reference~\cite{nosaka2008gigantic}.
By contrast, $|s_{1133}/d_{\mathrm{EDL}}|$ derived from $L_2$ is invariant to a systematic $V_{\mathrm{WE}}$ offset and is therefore unaffected by the PZC uncertainty, as it is extracted from the slope.
The five reverse-sweep points included at $V_{\mathrm{pp}} = 0.75$\,V agree with the forward-sweep points within error, and no appreciable hysteresis or drift was observed within the measurement time window.
An $F$-test against an extended model that adds a quadratic term $V_{\mathrm{WE}}^2$ to Eq.~\eqref{eq:fit_equation} shows no significant improvement over the linear form, confirming that a description linear in $V_{\mathrm{WE}}$ is sufficient (see the supplementary material).

\begin{figure}[tbp]
  \centering
  \includegraphics[width=0.85\linewidth]{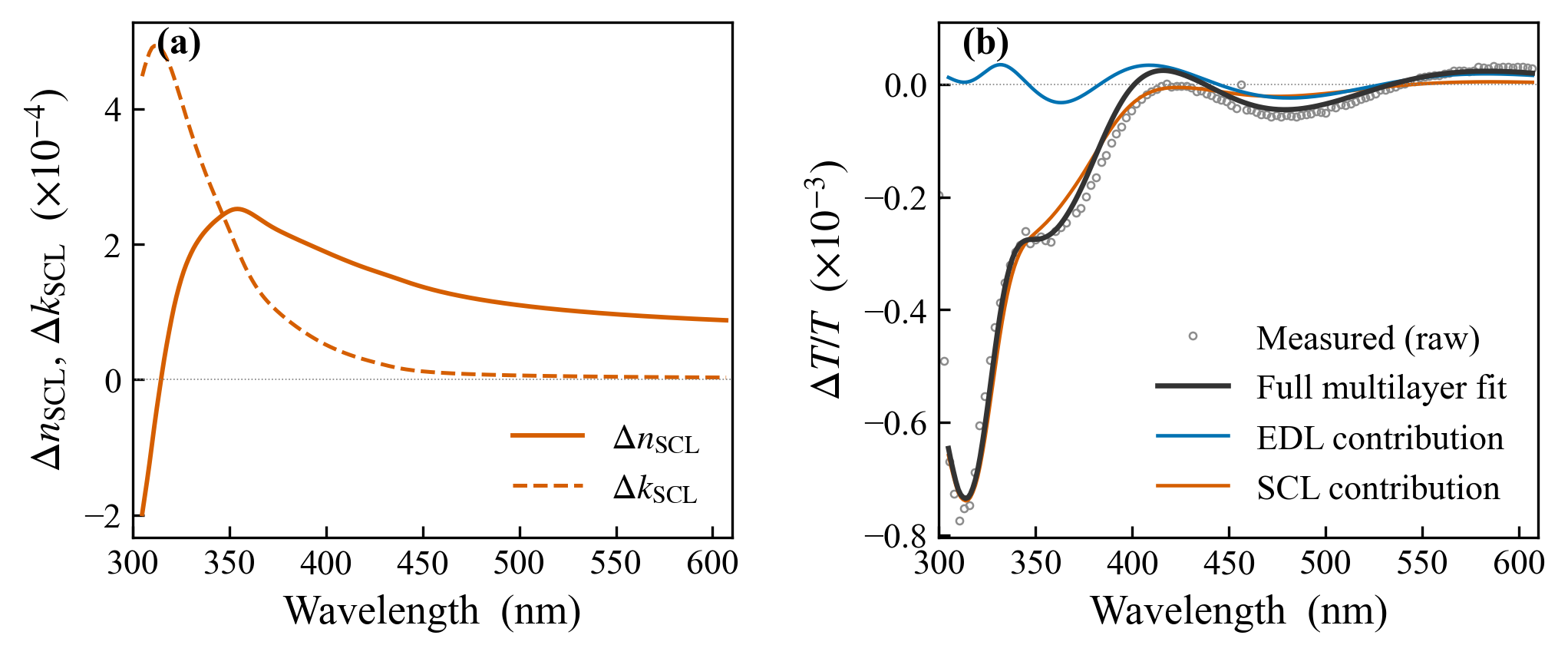}
  \caption{
    Simultaneous separation, within a single spectrum, of the broadband refractive-index change originating from the EDL and the Burstein--Moss-type response of the ITO space-charge layer (SCL), obtained by the multilayer transfer-matrix fit.
    The measured $\Delta T/T$ spectrum (raw data) is overlaid with the full multilayer fit (Full) together with its EDL-only and SCL-only components, demonstrating that the two can be separately extracted within the adopted SCL model from the difference in their spectral shapes.
  }
  \label{fig:separability}
\end{figure}

Over the spectral region accessed by the broadband white-light probe, the broadband refractive-index change originating from the EDL and the Burstein--Moss-type band-edge-shift response corresponding to the electron-density change of the ITO space-charge layer (SCL) can be separated within a single spectrum from the difference in their spectral shapes (multilayer transfer-matrix fit, Fig.~\ref{fig:separability}).
This fit adopts an SCL model in which the absorption change arising from the Burstein--Moss (band-population) effect in the ITO SCL is converted into a refractive-index change through a multi-oscillator Lorentz model that satisfies the Kramers--Kronig relations, the same treatment used by Nosaka \textit{et al.}~\cite{nosaka2008gigantic} and Okada \textit{et al.}~\cite{okada2022interfacial}.

Using $\eta_{\mathrm{AC}} = 0.99$, $n = 1.33$, and $\eta_{\mathrm{DC}} = 1$, and mapping $r_{13}$ and $s_{1133}/d_{\mathrm{EDL}}$ separately from the intercept $L_1$ and slope $L_2$ of the same WLS fit, we obtain
\begin{align}
  |r_{13}| &= (1.18 \pm 0.06_{\mathrm{PZC}})\times10^{2}\;\text{pm/V},
  \label{eq:r_result}\\
  \left|\frac{s_{1133}}{d_{\mathrm{EDL}}}\right|
  &= 33.0 \pm 5.6\;\text{pm/V}^{2}.
  \label{eq:s_result}
\end{align}
Because the fit extracts the EDL and SCL components within the same spectrum, the regression fixes the magnitudes $|L_1|,|L_2|$ and their relative sign, while the overall sign of the plotted $\Delta L_{\mathrm{opt}}$ is a measurement-phase convention that is not used.
Taking the established SCL response, a negative refractive-index change in the visible under positive bias~\cite{nosaka2008gigantic}, as the absolute reference, the co-fit places the EDL component in the same sense, so $\Delta n_{\mathrm{water}} < 0$ under positive bias and, from $r_{13} = -2\Delta n_{\mathrm{water}}/(n^{3} E)$, $r_{13} > 0$; the relative sign $L_2/L_1 < 0$ then fixes $s_{1133} < 0$.
The uncertainties quoted for $|r_{13}|$ and $|s_{1133}/d_{\mathrm{EDL}}|$ propagate from those of $L_1$ and $L_2$, respectively.
$|s_{1133}/d_{\mathrm{EDL}}|$ is, however, conditional on the choice $\eta_{\mathrm{DC}} = 1$; because $\eta_{\mathrm{AC}}, \eta_{\mathrm{DC}} \le 1$ holds physically, it should be read as a lower bound, as should $|r_{13}|$.

Our $|r_{13}|$ agrees at the order-of-magnitude level with the value of $165 \pm 17$\,pm/V reported by Okada \textit{et al.}~\cite{okada2022interfacial} for a nominally identical cell, lying about 29\,\% below it. Both studies operate without PZC calibration and under the same effective electro-optic coefficient definition, so this offset combines differences in measurement configuration, fit strategy, and cell-to-cell systematics, and cannot be attributed to any one of these alone.

Converting the extracted $|s_{1133}/d_{\mathrm{EDL}}|$ to an absolute susceptibility $\chi^{(3)}_{1133}(\omega;\omega,0,0)$ requires the relation $|\chi^{(3),\mathrm{int}}_{1133}| = (n^4/3)\,|s_{1133}|$, derived from the electro-optic indicatrix expansion of Boyd~\cite{boyd2020}, together with an assumed value of $d_{\mathrm{EDL}}$. Here $\chi^{(3),\mathrm{int}}_{1133}$ is a layer-effective, bulk-like susceptibility (m$^2$/V$^2$) obtained by treating the EDL of thickness $d_{\mathrm{EDL}}$ as a uniform-field layer. The static-field-induced third-order term that SHG/SFG studies separate from the interfacial $\chi^{(2)}$~\cite{gonella2016second} shares this dimension but describes a field-induced sum-frequency process $\chi^{(3)}(\omega_1+\omega_2;\omega_1,\omega_2,0)$, distinct in its frequency arguments from the DC Kerr response $\chi^{(3)}(\omega;\omega,0,0)$ probed here, so absolute magnitudes are not directly comparable.
Substituting physically reasonable $d_{\mathrm{EDL}}$ scenarios (a Stern-layer-dominated $\sim 0.6\,\mathrm{nm}$~\cite{brown2016effect}, the Debye length alone $\approx 1.0\,\mathrm{nm}$, and Stern $+$ Debye $\sim 1.6\,\mathrm{nm}$) gives $|\chi^{(3),\mathrm{int}}_{1133}| = 2.1\times10^{-20}$, $3.5\times10^{-20}$, and $5.5\times10^{-20}\,\mathrm{m^2/V^2}$, respectively. Relative to the visible-range bulk-water Kerr value $|\chi^{(3)}_{\mathrm{bulk}}| \approx 5.5\times10^{-21}\,\mathrm{m^2/V^2}$ (converted from the bulk Kerr constant $B$; see the supplementary material)~\cite{aroney1976kerr,zahn1985electro}, these correspond to ratios of about $4\times$, $6\times$, and $10\times$; even the lowest-end scenario exceeds the bulk value by several-fold.

The field-quadratic term of the present index response is the interfacial DC
Kerr coefficient $s_{1133}$. $\Delta L_{\mathrm{opt}}/R_p$ is linear in
$V_{\mathrm{WE}}$, and this linear structure is reproduced across independent
days and reversed-electrode cells (supplementary material); no field dependence
of $s_{1133}$ is detected over the measured range, so we treat it as a
well-defined third-order coefficient. Its magnitude exceeds the bulk-water DC
Kerr value~\cite{aroney1976kerr,zahn1985electro} several-fold, and, with the SCL
refractive-index change taken as the absolute reference used above, it is of
opposite sign.

Microscopically, the interfacial DC Kerr coefficient $s_{1133}$ separates
into three molecular contributions: an electronic (distortion)
hyperpolarizability term, a dipole--hyperpolarizability ($\mu\beta$) term,
and a permanent-dipole orientation term~\cite{orttung1963kerr,buckingham1955theoretical}.
The electronic term is nearly isotropic and environment-insensitive, and in
bulk water it accounts for only a few percent of the measured Kerr
constant~\cite{orttung1963kerr}; it can therefore produce neither the opposite
sign relative to bulk nor the above-bulk magnitude reported above, and is not
the origin of either. Both features thus reside in the two orientation-sensitive
terms---the $\mu\beta$ and orientation contributions---which a single-frequency
measurement cannot separate. Their distinct temperature ($T$) dependences (the
electronic term nearly constant, the $\mu\beta$ term scaling as $T^{-1}$, and
the orientation term as $T^{-2}$) offer an experimental route to disentangle
them~\cite{orttung1963kerr}, which we leave to future work. The present
single-frequency characterization at 21\,Hz therefore does not assign a
specific molecular mechanism, and we report the observable as the
layer-effective DC Kerr response of the charged water/ITO EDL.
The extracted coefficient is a property of the specific interface (substrate,
electrolyte, and solvent) rather than a constant intrinsic to bulk water. The
interfacial Pockels coefficient depends strongly on the electrode material, is
of opposite sign at the TiO$_2$ interface relative to ITO~\cite{tanimoto2023pockels},
and varies with the solvent~\cite{okada2022interfacial}; it is orders of
magnitude smaller at noble-metal interfaces such as Pt and Ag, which form no
oxide film, than at oxide electrodes~\cite{tanimoto2023pockels,yukita2018mechanisms}.

At a fixed DC bias the 1f response contains the field-quadratic mixing term
$2 s_{1133} E_{\mathrm{DC}}$, besides the bias-independent Pockels term.
Through $E_{\mathrm{DC}}$ this second term carries an electrolyte-concentration
dependence that the bare Pockels coefficient $r_{13}$ does not: a standing field
at the operating point scales with the Debye length as
$E_{\mathrm{DC}} \propto 1/d_{\mathrm{EDL}} \propto M^{1/2}$. This
provides a candidate origin for the $M^{1/2}$ concentration dependence reported
for the interfacial EDL Pockels signal~\cite{tokunaga2007pockels,nosaka2008gigantic,yukita2018mechanisms}.
Testing it would require concentration-resolved measurements beyond the present
single-electrolyte ($0.1\,\mathrm{M}$) study, and is left for future work.

The present method and the $\chi^{(2)}_{\mathrm{eff}}$ separation framework of
SHG/SFG differ in both the frequency arguments ($\omega$ versus $2\omega$) and
the DC-field input (the local field $E_{\mathrm{DC}}$ [V/m] versus the
interfacial potential $\Phi_0$ [V]). Although
$\chi^{(3)}(\omega;\omega,0,0)$ and $\chi^{(3)}(2\omega;\omega,\omega,0)$ are the
same third-order susceptibility, the two mixing processes weight different
molecular components through the orientational averaging, so the bulk-water
$\chi^{(3)}_{\mathrm{b}}$ invariance convention adopted in the SHG/SFG
literature~\cite{ohno2019beyond,dalstein2019direct}---an argument made under the
$2\omega$ arguments---cannot be transplanted to the DC Kerr $\chi^{(3)}$ here. A
direct comparison with absolute SHG/SFG values therefore calls for caution; we
instead anchor along the $\omega$ path on order-level consistency with
bulk-water Kerr constants at the same frequency
arguments~\cite{zahn1985electro,aroney1976kerr}.

Non-pure-capacitive contributions are bounded phenomenologically to
$\le 2.4\,\%$ of the observed slope $|L_{2}|$, i.e.\ $\le 1.85$\,pm/V$^{2}$, which is
not an appreciable systematic error on the reported
$|s_{1133}/d_{\mathrm{EDL}}|$ (see the supplementary material). A separate caveat concerns the sign: the EDL/SCL decomposition, and with it the sign assignment of the EDL response, is anchored to the adopted Burstein--Moss SCL model family~\cite{nosaka2008gigantic,okada2022interfacial} and is therefore not model-independent.

Using a 1f bias-mixing EO path, we simultaneously extract the Pockels and DC Kerr terms of the water/ITO interface from a single AC$+$DC sweep, obtaining $|r_{13}| = (1.18 \pm 0.06_{\mathrm{PZC}})\times10^{2}\,\mathrm{pm/V}$ and $|s_{1133}/d_{\mathrm{EDL}}| = 33.0 \pm 5.6\;\mathrm{pm/V^2}$ as lower bounds under $\eta_{\mathrm{AC}}, \eta_{\mathrm{DC}} \le 1$. The former agrees at the order-of-magnitude level with reported interfacial Pockels coefficients~\cite{tokunaga2007pockels,okada2022interfacial}, while the interfacial $|\chi^{(3),\mathrm{int}}_{1133}|$ lies several times above the visible-range DC Kerr response of bulk water evaluated at the same frequency arguments (numerical details above and in the supplementary material).
This observable is characterized as the layer-effective DC Kerr response of the charged water/ITO EDL, without assigning a specific microscopic mechanism. Its value is a property of the specific interface, tunable through the choice of electrode, electrolyte, and solvent rather than intrinsic to bulk water. Operating on the $\omega$ path, the method directly probes $\chi^{(3)}(\omega;\omega,0,0)$ and provides an electro-optic route complementary to SHG/SFG (the $2\omega$ path); independent determination of $d_{\mathrm{EDL}}$ and mapping of the frequency dispersion remain for future work.

\begin{acknowledgments}
This research was funded by a Grant-in-Aid for Scientific Research (B) (Grant Number JP25K01698) from the Japan Society for the Promotion of Science (JSPS).
\end{acknowledgments}

\section*{Author Declarations}
\subsection*{Conflict of Interest}
The authors have no conflicts to disclose.

\subsection*{Author Contributions}
\textbf{Soichiro Ashikaga}: Conceptualization (lead); Methodology (lead); Investigation (lead); Formal analysis (lead); Visualization (lead); Writing -- original draft (lead); Writing -- review \& editing (equal).

\textbf{Kazuaki Nakata}: Conceptualization (supporting); Methodology (supporting); Writing -- review \& editing (equal).

\textbf{Akihiro Okada}: Methodology (supporting); Formal analysis (supporting); Software (supporting); Writing -- review \& editing (equal).

\textbf{Takumi Takayanagi}: Methodology (supporting); Writing -- review \& editing (equal).

\textbf{Kyohei Yamashita}: Supervision (supporting); Writing -- review \& editing (equal).

\textbf{Takayoshi Kobayashi}: Supervision (supporting); Writing -- review \& editing (equal).

\textbf{Eiji Tokunaga}: Conceptualization (supporting); Resources (lead); Supervision (lead); Project administration (lead); Funding acquisition (lead); Writing -- review \& editing (equal).
\section*{Data Availability}
The data that support the findings of this study are available from the corresponding author upon reasonable request.
\clearpage
\section*{Supplementary Material}
\section{Apparatus geometry of the electrochemical cell and optical system}
\label{sec:si_geometry}

The electrochemical cell is a parallel-plate three-electrode configuration with
an ITO transparent electrode (300\,nm thick, on a quartz substrate) as the
working electrode (WE), together with a counter electrode (CE) of the same ITO
and an Ag/AgCl reference electrode (RE) in a 0.1\,M NaCl aqueous solution.
The light beam ($\sim 2\,\mathrm{mm}$ in diameter) is incident normally near the
center of the electrode.
The main apparatus parameters are summarized in Table~\ref{tab:si_geometry}.

\begin{table}[tbp]
  \centering
  \renewcommand{\arraystretch}{1.3}
  \caption{Main apparatus parameters of the electrochemical cell and optical system.}
  \label{tab:si_geometry}
  \begin{tabular}{@{}p{3.3cm}@{\hspace{1.8em}}p{5.9cm}@{\hspace{1.8em}}p{3.7cm}@{}}
    \toprule
    Item & Value & Notes \\
    \midrule
    Working electrode (WE) & ITO transparent electrode, 300\,nm thick, on quartz substrate & — \\
    Counter electrode (CE)  & ITO electrode of the same type & parallel-plate arrangement \\
    Reference electrode (RE) & Ag/AgCl & near WE, $< 5$\,mm \\
    Electrode gap $d_{\mathrm{cell}}$ & 17\,mm & WE--CE \\
    Effective wetted area & $1.1 \times 1\,\mathrm{cm}^2$ & — \\
    Electrolyte & 0.1\,M NaCl aqueous solution & — \\
    Light source & Xe lamp (white light, collimated and normally incident) & — \\
    Probe beam diameter & $\sim 2\,\mathrm{mm}$ (diameter) & near electrode center \\
    AC drive frequency & 21\,Hz & lock-in reference frequency \\
    AC drive amplitude $V_{\mathrm{pp}}$ (four groups) & 0.25, 0.50, 0.75, 1.00\,V & — \\
    Impedance buffer gain & 1\,V/V (unity gain) & for RE output readout \\
    Buffer input bias current & $\lesssim 50\,\mathrm{fA}$ & — \\
    \bottomrule
  \end{tabular}
\end{table}

\section{Broadband EIS and equivalent-circuit fitting}
\label{sec:si03}

To accurately evaluate the AC voltage amplitude applied to the electric double
layer (EDL), we acquired broadband electrochemical impedance spectroscopy (EIS)
on this cell.
The measurement was performed between the working electrode (ITO) and the
reference electrode (Ag/AgCl, located $\sim 5$\,mm from the working electrode)
over a frequency range from $63\,\mathrm{kHz}$ to $0.4$\,Hz, with an AC amplitude
of 50\,mV~rms, in a 0.1\,M NaCl electrolyte.
An auxiliary sweep extending down to $\sim 0.05$\,Hz was also acquired, but
because reproducibility degrades at low frequency from $1/f$ noise, only points
at or above $0.4$\,Hz were used in this fit.

\begin{figure}[tbp]
  \centering
  \includegraphics[width=0.7\linewidth]{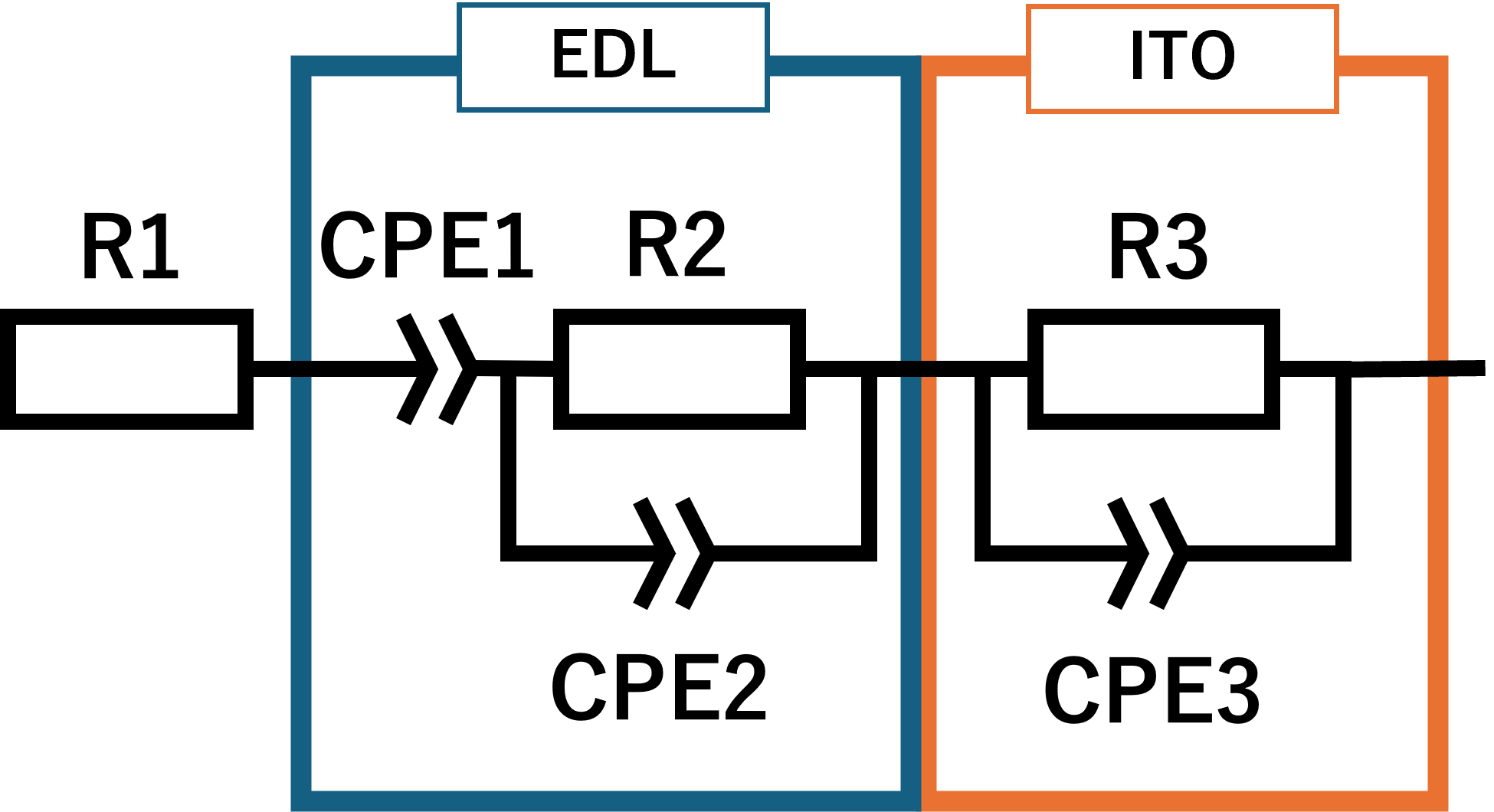}
  \caption{Three-block series equivalent circuit used in this work.
    $R_{1}$ is the bulk solution resistance;
    the EDL block (blue box) is the series combination of $\mathrm{CPE}_{1}$ and $R_{2}\,\|\,\mathrm{CPE}_{2}$,
    and the ITO space-charge layer (SCL) block (orange box) is represented by $R_{3}\,\|\,\mathrm{CPE}_{3}$.
    $\mathrm{CPE}_{n}$ is a constant phase element (non-ideal capacitance) and $R_{n}$ a parallel resistance.
    The equivalent-circuit parameters are determined by a free fit to the broadband EIS of this cell ($63$\,kHz--$0.4$\,Hz)
    (Table~\ref{tab:si03_m3_params}).
    Alt text: Circuit diagram of a three-block series network---a bulk solution
    resistance in series with an electric-double-layer block (a constant phase
    element followed by a resistor in parallel with a constant phase element) and
    an ITO space-charge-layer block (a resistor in parallel with a constant phase
    element).}
  \label{fig:si_equivalent_circuit}
\end{figure}

\begin{figure}[tbp]
  \centering
  \includegraphics[width=0.95\linewidth]{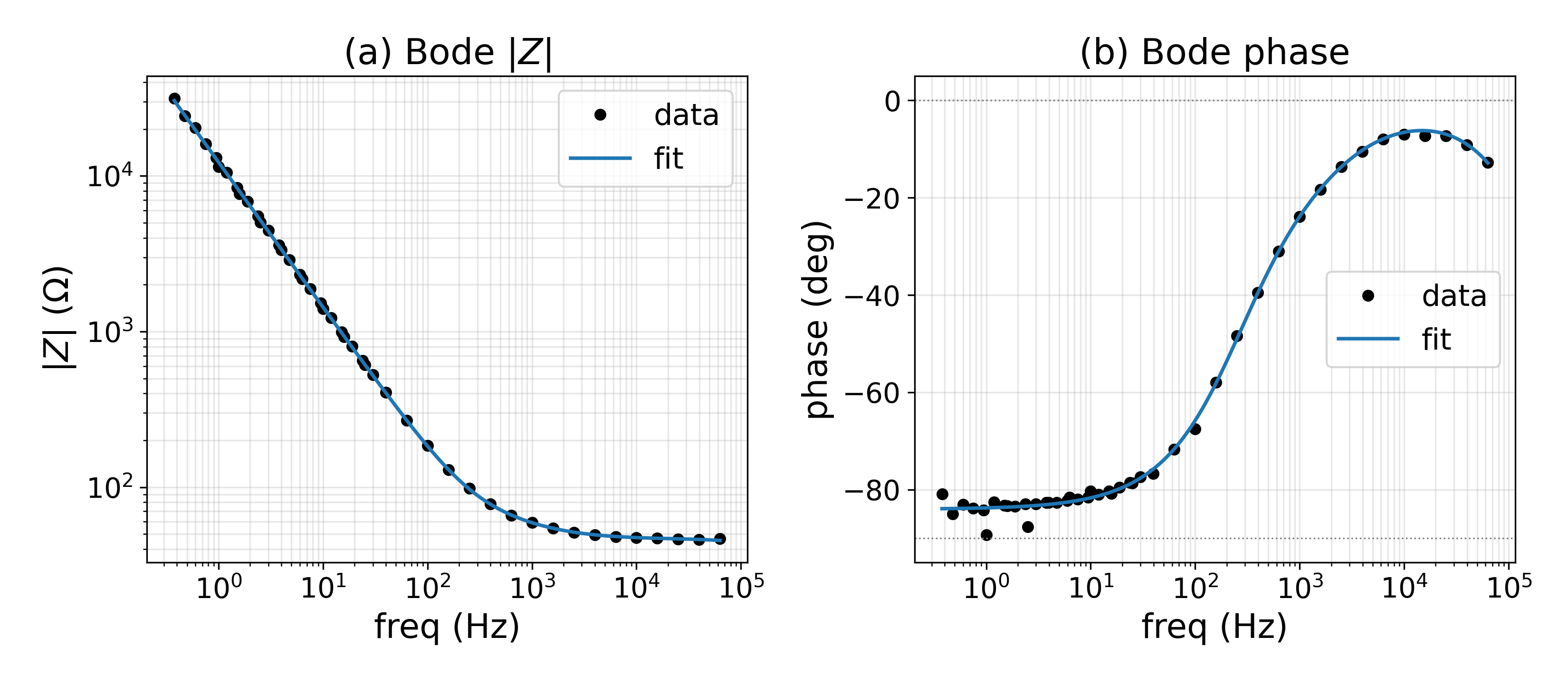}
  \caption{Bode plot of the broadband EIS.
    (a) The impedance magnitude $|Z|$ and (b) the phase angle are shown as functions of frequency.
    Black circles are the measured data ($0.4$\,Hz--$63$\,kHz),
    and the blue solid line is the result of fitting the equivalent circuit of
    Fig.~\ref{fig:si_equivalent_circuit} with all parameters free.
    This equivalent-circuit model reproduces both $|Z|$ and the phase over the entire measurement frequency window.
    Alt text: Two-panel Bode plot versus frequency from $0.4$\,Hz to $63$\,kHz;
    panel (a) shows the impedance magnitude and panel (b) the phase angle, with
    measured data as black circles and the equivalent-circuit fit as a blue line
    that overlaps the data across the band.}
  \label{fig:si_eis_bode}
\end{figure}

The equivalent circuit adopts a three-block series topology of the same form as
Okada \textit{et al.}~\cite{okada2022interfacial},
$R_{1}\,$--$\,\mathrm{CPE}_{1}\,$--$\,(R_{2}\,\|\,\mathrm{CPE}_{2})\,$--$\,(R_{3}\,\|\,\mathrm{CPE}_{3})$
(Fig.~\ref{fig:si_equivalent_circuit}), and the nine equivalent-circuit
parameters are re-determined by a free fit to the broadband EIS of this cell.
Here $\mathrm{CPE}_{n}$ denotes a constant phase element (non-ideal capacitance)
and $R_{n}$ a parallel resistance.
The resulting fit reproduces the data over the full band from $0.4$\,Hz to
$63$\,kHz, as shown in Fig.~\ref{fig:si_eis_bode}.

\begin{table}[tbp]
  \centering
  \caption{Results of the free fit of the equivalent circuit for this cell and
    $\eta_{\mathrm{AC}}(21\,\mathrm{Hz})$. $R_{i}$ is in $\Omega$ and
    $Q_{i}$ in F$\cdot$s$^{\alpha_{i}-1}$.}
  \label{tab:si03_m3_params}
  \begin{tabular}{cccccccccc}
    \toprule
    $R_{1}$ & $Q_{1}$ & $\alpha_{1}$ & $R_{2}$ & $Q_{2}$ & $\alpha_{2}$ & $R_{3}$ & $Q_{3}$ & $\alpha_{3}$ & $\eta_{\mathrm{AC}}$(21\,Hz) \\
    \midrule
    1.0 & $1.5{\times}10^{-5}$ & 0.93 & 13.1 & $1.0{\times}10^{-4}$ & 0.77 & 45.1 & $1.2{\times}10^{-8}$ & 1.00 & 0.991 \\
    \bottomrule
  \end{tabular}
\end{table}

The AC voltage amplitude partitioning ratio $\eta_{\mathrm{AC}}$ at $21$\,Hz is
given by the vector formula for the series impedance of the EDL block
$Z_{\mathrm{EDL}} = Z_{\mathrm{CPE}_{1}} + Z_{R_{2}\,\|\,\mathrm{CPE}_{2}}$ and the
ITO SCL block $Z_{\mathrm{ITO}} = Z_{R_{3}\,\|\,\mathrm{CPE}_{3}}$,
\begin{equation}
  \eta_{\mathrm{AC}}(21\,\mathrm{Hz})
  = \frac{|Z_{\mathrm{EDL}}(21\,\mathrm{Hz})|}{|Z_{\mathrm{EDL}}(21\,\mathrm{Hz}) + Z_{\mathrm{ITO}}(21\,\mathrm{Hz})|}
  \label{eq:si_eta_edl_divider}
\end{equation}
Substituting the equivalent-circuit fit values into
Eq.~\eqref{eq:si_eta_edl_divider} gives $\eta_{\mathrm{AC}}(21\,\mathrm{Hz}) = 0.991$
(the main-text headline adopts $\eta_{\mathrm{AC}} = 0.99$, rounded to two decimal places).

Because the reference electrode is placed a few mm from the working electrode in
the three-electrode configuration, the bulk solution resistance $R_{1}$ is small
and is pinned at the lower bound of $1\,\Omega$ in the fit, so it is not
identified independently. However, the EDL and ITO SCL block impedances are much
larger than $R_{1}$, so the AC voltage amplitude partitioning ratio
$\eta_{\mathrm{AC}}(21\,\mathrm{Hz}) = |Z_{\mathrm{EDL}}|/|Z_{\mathrm{EDL}}+Z_{\mathrm{ITO}}|$
is set by these two blocks and is insensitive to $R_{1}$.

\section{Working assumption for the stability of \texorpdfstring{$\eta_{\mathrm{AC}}$}{eta\_AC} from DC-bias-dependent EIS}
\label{sec:si04}

In this work, $\eta_{\mathrm{AC}} = 0.99$ is determined from broadband EIS taken at a single bias point at open-circuit potential, and it is applied across the full bias range of the optical measurements ($V_{\mathrm{WE}} \approx 0$--$+0.8$\,V).
This is a working assumption rather than a strict per-bias verification, supported by two observations: across the five DC-bias-dependent EIS spectra acquired over the effective bias range of the optical measurements, the $21$\,Hz phase stays within a $2.4^{\circ}$ window from $-69.8^{\circ}$ to $-72.2^{\circ}$ with no clear bias dependence, and the $|Z|$ Bode curves shift monotonically between biases while preserving their shape, indicating that the equivalent-circuit topology is conserved over this bias range.

Directly verifying the bias dependence of the partitioning ratio $\eta_{\mathrm{AC}}$ is left for future work; in this work, $\eta_{\mathrm{AC}} = 0.99$ is applied across the full bias range under the working assumption based on the two observations above.

\section{Robustness of the regression statistics}
\label{sec:si05}

The linear fit of the main text is $\Delta L_{\mathrm{opt}}/R_p = L_{1} + L_{2}\,V_{\mathrm{WE}}$ (Eq.~\eqref{eq:fit_equation}), with intercept $L_{1}$ and slope $L_{2}$.

Throughout the main text and this supplementary material, the coefficient uncertainties are the $1\sigma$ values from the unscaled WLS formal covariance, weighted by $1/\sigma^{2}$ with the conservative per-point refractive-index uncertainties from the multilayer transfer-matrix fit.

An F-test assessed the significance of an extended model (linear+quadratic) that adds a quadratic term $L_{2}^{(2)}\,V_{\mathrm{WE}}^{2}$ to the linear model $L_{1} + L_{2}\,V_{\mathrm{WE}}$; this extension corresponds to adding a cubic term $tE^{3}$ to the $\Delta n$--$E$ relation. The F statistic follows an $F(1, N-3)$ distribution (degrees of freedom $N-2 = 31$ versus $N-3 = 30$), and the smallest $p$ value across the four independent $V_{\mathrm{pp}}$-group fits and the 33-point combined fit is $0.19$. The quadratic term provides no significant improvement over the linear model, which is the statistical basis for the main-text statement that the linear (first-order in $V_{\mathrm{WE}}$) fit is sufficient. Given the $N = 33$ points and the conservative weighting, the statistical power to detect a quadratic term is limited, and within this power no quadratic term was detected.

Beyond the regression statistics, the slope $L_{2}$ admits a depth-integral interpretation. Equation~\eqref{eq:fit_equation} is a reduction in which the EDL field profiles $E_{\mathrm{AC}}(z)$ and $E_{\mathrm{DC}}(z)$ are taken as uniform; in general the slope is an effective coefficient corresponding to the depth-weighted integral
\begin{equation}
  L_{2}^{(\mathrm{general})}
  \propto \int_{0}^{d_{\mathrm{EDL}}} s_{1133}(z)\,E_{\mathrm{AC}}(z)\,E_{\mathrm{DC}}(z)\,dz
  \label{eq:si05_general}
\end{equation}
Setting the Stern--diffuse field profiles $E_{\mathrm{AC,DC}}(z)$ uniform within the EDL ($E \approx \mathrm{const.} \times V/d_{\mathrm{EDL}}$) reduces the integral to the uniform-layer expression of the main text. Refinement of the depth profile (separation of the Stern and diffuse layers, and the inhomogeneity of $s_{1133}(z)$) is left for future work.

\section{Reproducibility across independent days and electrodes, and propagation of the PZC uncertainty}
\label{sec:si07}

\begin{figure}[tbp]
  \centering
  \includegraphics[width=0.78\linewidth]{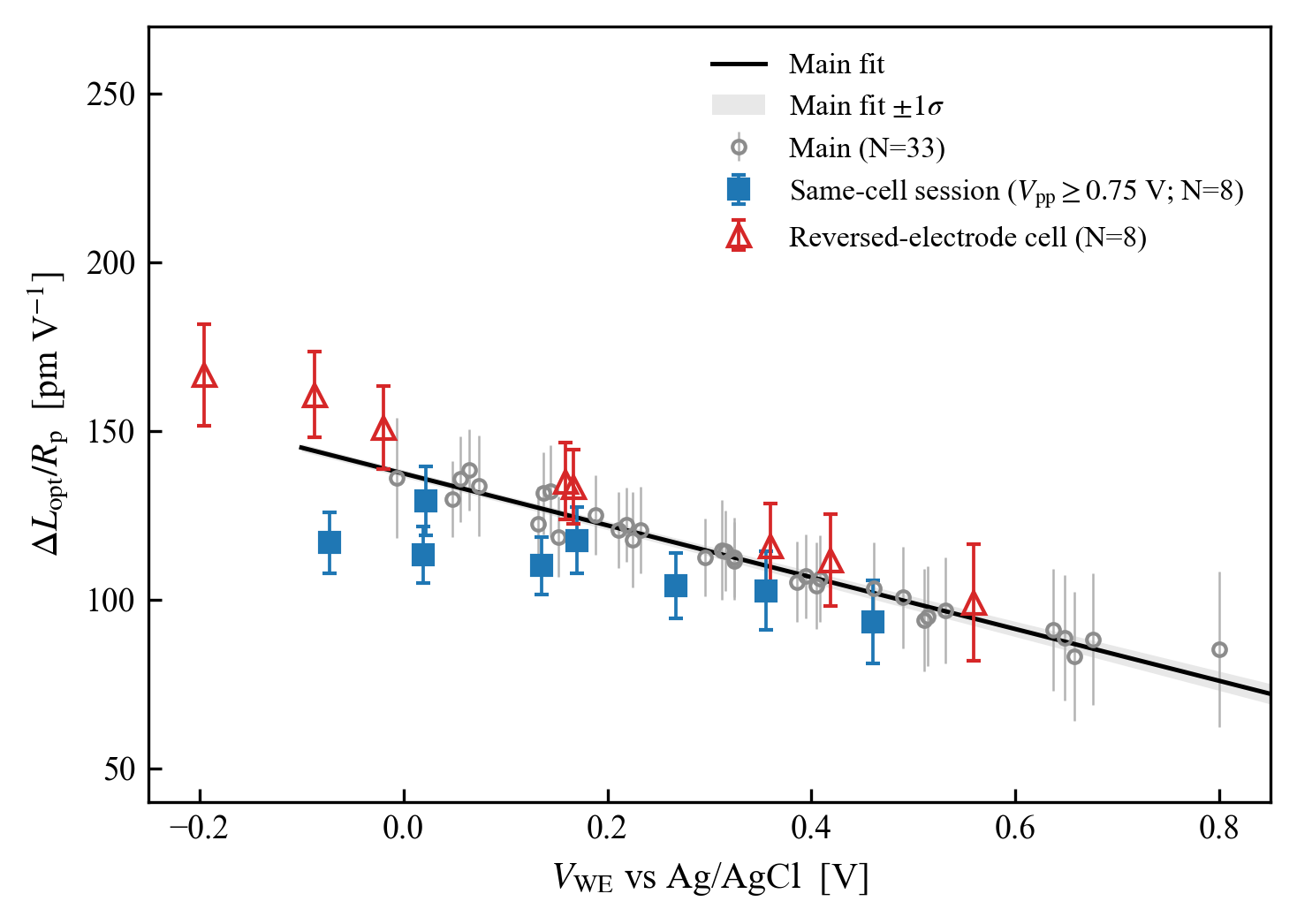}
  \caption{Auxiliary comparison of $\Delta L_{\mathrm{opt}}/R_{p}$ versus $V_{\mathrm{WE}}$ across independent days and electrodes (consistency within the present auxiliary uncertainties).
    Gray circles, the main dataset ($N = 33$); black line, the main-dataset fit ($\pm 1\sigma$ band).
    Blue squares, the auxiliary same-cell session (8 points over $V_{\mathrm{pp}} = 0.75$--$1.00$\,V);
    open red triangles, the reversed-electrode cell (a separate cell with the working- and counter-electrode roles swapped,
    8 points over $V_{\mathrm{pp}} = 0.75$--$1.00$\,V).
    The individual intercepts of the two auxiliary datasets bracket the main-dataset fit
    (per-point mean offset $-1.8 \pm 3.4$\,pm V$^{-1}$),
    and the combined 16-point WLS fit lies within $0.5\sigma$ in slope and $1.26\sigma$ low in intercept.
    Alt text: Scatter plot of the normalized optical path-length change versus
    working-electrode bias, comparing the 33-point main dataset (gray circles with
    a black fit line and a $\pm 1\sigma$ band) against an auxiliary same-cell
    session (blue squares) and a reversed-electrode cell (open red triangles), all
    following the same linear trend.}
  \label{fig:si_cross_day}
\end{figure}

To verify independently that the main dataset ($N = 33$) does not rest on a single session, the same weighted least-squares linear fit was applied to two auxiliary datasets: an auxiliary same-cell session (a separate-day session on the same cell as the main dataset, $N = 8$ over $V_{\mathrm{pp}} = 0.75$--$1.00$\,V), and a reversed-electrode cell (a separate cell with the working- and counter-electrode roles swapped, $N = 8$ over the same $V_{\mathrm{pp}}$ range). Their intercepts and slopes were compared with the main dataset (Fig.~\ref{fig:si_cross_day}).
The individual intercepts of the two auxiliary datasets bracket the main-dataset fit (auxiliary session $-2.99\sigma$, reversed-electrode $+1.76\sigma$); the combined 16-point WLS fit agrees with the main-dataset value in slope to within $0.46\sigma$, while its intercept lies $-1.26\sigma$ low.

The cross-day shift of the intercept is consistent in order of magnitude with the day-to-day shift of $V_{\mathrm{WE}}$ at $V_{\mathrm{app,DC}} = 0$ ($\sim 0.15$--$0.18$\,V), and is read as a systematic difference that can include reference-electrode liquid-junction-potential drift and surface-state renewal. The main result is observed as a linear $\Delta L_{\mathrm{opt}}/R_{p}$ versus $V_{\mathrm{WE}}$ structure across independent days and independent electrode configurations, whereas the quantitative reproducibility of the coefficient values across independent cells remains to be verified.

Next, we analyze how the point of zero charge (PZC) uncertainty propagates. Because the PZC is not measured independently in this study, the horizontal axis of the measurement is the PZC-uncorrected $V_{\mathrm{WE}}$ (vs Ag/AgCl). The PZC uncertainty propagates to $r_{13}$ through the linear model $\Delta L_{\mathrm{opt}}/R_p = L_{1} + L_{2}\,V_{\mathrm{WE}}$ of Eq.~\eqref{eq:fit_equation}: under a rigid horizontal re-zeroing $\Delta L_{\mathrm{opt}}/R_p = L_{1}' + L_{2}\,(V_{\mathrm{WE}} - V_{\mathrm{PZC}})$, the slope $L_{2}$ is invariant while the intercept shifts as $L_{1} = L_{1}' - L_{2}\,V_{\mathrm{PZC}}$. Hence $\sigma_{L_{1}}^{(\mathrm{PZC})} = |L_{2}|\,\sigma_{\mathrm{PZC}}$, and the coefficient propagating this uncertainty into $|r_{13}| = 2|L_{1}|/(\eta_{\mathrm{AC}}\,n^{3})$ is $2|L_{2}|/(\eta_{\mathrm{AC}}\,n^{3})$.

To bound the PZC contribution to $r_{13}$, we adopt the Bessel-corrected sample standard deviation of $V_{\mathrm{WE}}|_{V_{\mathrm{app,DC}} = 0}$ across the cross-day sessions, $\sigma_s \approx 0.092\,\mathrm{V}$ (obtained from the cross-day reproducibility analysis above), as a conservative upper-bound proxy for the PZC uncertainty:

\begin{equation}
  \sigma_{r_{13}}^{(\mathrm{PZC})}
  \approx \frac{2\,|L_2|\,\sigma_s}{\eta_{\mathrm{AC}}\,n^3}
  = \frac{2 \times 76.9 \times 0.092}{0.99 \times 1.33^3}
  \approx 6.07\,\mathrm{pm/V}
  \approx 0.06 \times 10^2\,\mathrm{pm/V}.
  \label{eq:si07_pzc_r13}
\end{equation}

This provides the basis for the PZC systematic term $0.06 \times 10^2\,\mathrm{pm/V}$ in the main-text value $|r_{13}| = (1.18 \pm 0.06_{\mathrm{PZC}}) \times 10^2\,\mathrm{pm/V}$, where the statistical component is carried separately by the $L_1$ uncertainty of the weighted least-squares fit. Because the two same-day sessions entering $\sigma_s$ are strongly correlated, the effective degrees of freedom are $\le 3$, and the value is adopted conservatively as an inflated upper bound. The origin of the PZC shift (a mixture of PZC variation, drift of the reference-electrode liquid-junction potential over time, and renewal of the ITO surface state) is not resolved by the present data, so this value is taken as an upper-bound estimate of the PZC-derived contribution.

\section{Bounding of non-capacitive contributions to the 1f signal}
\label{sec:si_noncap}

Non-pure-capacitive contributions to the 1f signal---Faradaic rectification,
capacitive nonlinearity, and electrical noise---were bounded from the DC-bias
dependence of the 1f cell current measured at 21\,Hz on the reversed-electrode
cell ($V_{\mathrm{pp}} = 0.75$\,V, five points), together with an
endpoint-excluded fit of the main dataset. Assigning the entire fractional bias
variation of the 1f current to a spurious (non-electro-optic) origin gives a
phenomenological upper bound of $\le 2.4\,\%$ of the observed slope
$|L_2|$, i.e.\ $\le 1.85$\,pm/V$^2$. This is not an appreciable systematic on the
reported $|s_{1133}/d_{\mathrm{EDL}}| = 33.0 \pm 5.6$\,pm/V$^2$, and is
therefore not propagated into the error budget.

\section{Conversion to the bulk-water DC Kerr value (same frequency arguments)}
\label{sec:si_chi3conv}

To compare the interfacial value $|\chi^{(3),\mathrm{int}}_{1133}|$ reported in the
main text with the DC Kerr response of bulk water, both are evaluated at the same
frequency arguments $\chi^{(3)}(\omega;\omega,0,0)$. The conversion from the
bulk-water Kerr constant $B$ is given below.

The Kerr birefringence is defined by
\begin{equation}
  n_{\parallel}-n_{\perp} = \lambda B E^{2},
\end{equation}
where $n_{\parallel},n_{\perp}$ are the refractive indices for polarization parallel
and perpendicular to the DC field $z$, and $B$ is the Kerr constant
(units $\mathrm{m\,V^{-2}}$). At zero field $n_{\parallel}=n_{\perp}=n_{0}$, so writing
the field-induced change of each polarization as
$\Delta n_{i}=(3/2n_{0})\,\chi^{(3)}_{iizz}E^{2}$ gives
$n_{\parallel}-n_{\perp}=\Delta n_{\parallel}-\Delta n_{\perp}$ and
\begin{equation}
  B = \frac{3}{2 n_{0}\lambda}\bigl(\chi^{(3)}_{zzzz}-\chi^{(3)}_{xxzz}\bigr).
\end{equation}
The Kerr response of water is dominated by the temperature-dependent terms arising
from the permanent dipole (the $\mu\beta$ and orientation terms); the electronic
(distortion) term, which does not conserve the trace, accounts for only a few
percent of the measured value~\cite{orttung1963kerr} (see the main text). In this
orientation-dominated limit, reorientation conserves the trace of the polarizability
tensor, so that $\chi^{(3)}_{zzzz}=-2\,\chi^{(3)}_{xxzz}$, and
\begin{equation}
  \chi^{(3)}_{xxzz} = -\frac{2 n_{0}\lambda}{9}\,B .
  \label{eq:si_chi3bulk}
\end{equation}
This is the same tensor component ($xxzz=1133$) as the interfacial conversion
$\chi^{(3),\mathrm{int}}_{1133}=-(n^{4}/3)\,s_{1133}$ of the main text, so the two
are compared under the same normalization. The $\parallel/\perp$ partition
($\chi^{(3)}_{zzzz}=-2\,\chi^{(3)}_{xxzz}$) is the main model dependence of this
conversion; the finite deviation associated with the $\mu\beta$ term is absorbed
into the order-level comparison.

With $n_{0}=1.33$, the value $B=2.96\times10^{-14}\,\mathrm{m\,V^{-2}}$
(633\,nm~\cite{aroney1976kerr}) gives
$|\chi^{(3)}_{\mathrm{bulk}}|=5.5\times10^{-21}\,\mathrm{m^{2}\,V^{-2}}$, and
$B=3.4$--$3.6\times10^{-14}\,\mathrm{m\,V^{-2}}$ (590\,nm~\cite{zahn1985electro})
gives $\sim6\times10^{-21}\,\mathrm{m^{2}\,V^{-2}}$. We adopt the Aroney 633\,nm value
$5.5\times10^{-21}\,\mathrm{m^{2}\,V^{-2}}$ as the bulk reference. These are values
reported in SI units, consistent after SI conversion with the esu-reported original
data~\cite{orttung1963kerr}.

\bibliography{refs}

\end{document}